# The kinetics and modes of gold nanowire breaking

Honghai Liu[1], Enyong Jiang[1], Haili Bai[1], Ping Wu[1], Zhiqing Li[1], and Chang Q Sun[2*]

[1] Tianjin Key Laboratory of Low Dimensional Materials Physics and Preparing Technology and Institute of Advanced Materials Physics, Faculty of Science, Tianjin University, Tianjin 300072, People's Republic of China

[2] School of Electrical and Electronic Engineering, Nanyang Technological University, Singapore 639798

**Abstract**

Molecular dynamics calculations revealed that the temperature of operation and the applied tensile force ($f$) determine not only the kinetics but also the mode and duration of Au nanowire breaking. In the tensile force range of 0.018 and 0.1 nN/atom, structure transformation of the wire occurs prior to breaking at random positions. The gold wire breaks abruptly when the $f$ is stronger than 0.1nN/atom but no rupture occurs at all when the $f$ is weaker than 0.018 nN/atom. At higher temperatures and under stronger tensile forces, the wire breaks even faster.

Kewywords: Nanostructures; rupture; molecular dynamics; stability

* Authers to whom all correspondence should be addressed:
E.Y. Jiang, Faculty of Science, Tianjin University, Tianjin 300072, China; Electronic Address: eyjiang@tju.edu.cn.
C.Q.Sun, school of Electrical and Electronic Engineering, Nanyang Technological University, Singapore 639798; Fax: 65 6793 3319; Electronic address: ecqsun@ntu.edu.sg

## 1. Introduction

Metallic nanowires have attracted considerable interest due to their novel properties and potential applications to upcoming technologies such as nanoelectronic and mechanical devices [1,2,3,4,5,6,7,8,9,10]. One of the most significant issues in nanowire applications is the structural stability of the wires under various conditions such as chemical, mechanical, and thermal stimuli [11,12,13]. Generally, the extension of a nanowire is subject to two effects when the wire is under tension. The wire is stretched and prolonged either in a constant speed under a non-constant tensile force or in different speed under a constant tensile load (e.g. hanging a heavy body under a piece of the wire). However, the kinetics, the modes, and the breaking times of the wire breaking under different conditions are poorly known.

Normally, the structural and thermal stability of a nanowire is studied by pulling one or both ends of the wire with the wire being prolonged in a constant speed [14,15,16,17,18,19]. However, a nanowire, in practice, is stretched by a constant tensile force [20,21]. Therefore, applying a constant tensile force is more reasonable, which has been seldom studied. In the present work, we report our findings in examining the possible breaking mechanism of a gold nanowire by stretching it with constant tensile forces. Results show that the temperature of operating and the strength of the tensile force dominate the kinetics, the mode, and the lifetime of gold wire breaking. At higher temperatures and under stronger tensile forces the wire will break in a shorter time. The nanowire breaks in different ways subjecting to the joint effect of the temperatures of operation and the applied stretching forces.

## 2. Principles and methods

The interaction between the Au atoms in the nanowire is described by the many-body tight-binding potential [22],

$$E_i = -\left\{\sum_j \xi^2 \exp\left[-2q\left(\frac{r_{ij}}{r_0}-1\right)\right]\right\}^{1/2} + \sum_j A\exp\left[-p\left(\frac{r_{ij}}{r_0}-1\right)\right], \quad (1)$$

where $r_{ij}$ is the distance between atom $i$ and atom $j$, and $r_0$ is the first-neighbor distance at equilibrium. The parameters $A$, $p$, $q$ and $\xi$ are obtained from the experimental data concerning cohesive energy, lattice parameter, bulk and shear elastic modulus, respectively (Table I in Ref. 22). Then the force between the two atoms is,

$$F_{ij} = -\sum_{j\neq i}\left\{\frac{\partial E_i}{\partial r_{ij}} + \frac{\partial E_j}{\partial r_{ij}}\right\}, \quad (2)$$

It is known that the breaking of a wire means the breaking of the all chemical bonds on the sites surrounding the spots of broken [23]. When the bond rupture occurs, the interatomic force decreases rapidly. So the breaking of a chemical bond could be determined by the relation,

$$D = \partial F_{ij}/\partial r_{ij} = -\sum_{j\neq i}\left\{\frac{\partial^2 E_i}{\partial r_{ij}^2} + \frac{\partial^2 E_j}{\partial r_{ij}^2}\right\} = 0.$$

The inflection point of the $D$ curve should correspond to the maximal length of the chemical bond. Figure 1 illustrates the pair potential, the first (force) and the second derivative of the potential (curve $D$). The maximal of the D curve corresponds to the breaking point of the bond. Basing on this definition, we calculated the breaking distance of the Au-Au distance in the bulk to be around 3.2 Å at 50K, agreeing with the reported findings (Ref. 7 and the references therein).

We cut a piece of Au nanowire with a $0.816\times 0.816$ nm$^2$ or $4 \times 4$ atoms cross-section from Au FCC bulk along the <100> direction. After a relaxing process of 30 ps at 50K, the FCC structure is transformed to a BCT wire with rhombic cross section, and the edge size of the cross section becomes 0.979 nm, which reproduces the transformation described by Diao et al [11] The relaxed BCT sample will be used as the original structure in the following study.

Figure 2 illustrates the configuration for wire stretching. Exerting a constant force on the wire means that all atoms in the pulling region will share equally the entire force applied. In the present work, we exert a weak constant force on each atom uniformly in the pulling region, then the entire force F ($F = nf$, see Figure 2) on the wire will keep constant. For convenience, we study the force $f$ on a representative atom in the pulling region, instead of the $F$.

## 3. Results and discussion

First, we exert a continuous constant tensile force amounted at $f = 1$ nN/atom along the axis of the wire and change the temperature. Result in Figure 3a shows that the lifetime of nanowire before breaking becomes shorter under a relatively higher temperature. Next, by keeping the temperature at 50K, we changed the tensile force $f$ to see the life time of the wire. Results in Figure 3b show that the breaking time of the wire varies with the $f$ applied. A larger $f$ value will cause the wire to break in shorter time, e.g. when $f = 0.05$ nN/atom, the breaking time is 47.1 ps, while $f = 0.1$ nN/atom the breaking time turns to be 14.7 ps. It is interesting to note that the curve in Figure 3b could be divided into three regions: (a) at $f > 0.1$ nN/atom, the breaking time of the Au wire decreases slowly when the the force applied is increased; (b) when the force is between 0.018 and 0.1 nN/atom, the breaking time of the Au wire changes irregularly; (c) at $f < 0.018$ nN/atom, no breaking occurs to the Au wire even in longer time (140ps).

Since the breaking of the chemical bond is arisen from the separation of the concerned atoms, the breaking time of the chemical bond is determined by the separating speed of the atoms. Therefore, the velocity of the atoms moving along the axis of the wire relates to the breaking of the Au wire. To raise the temperature of operation means to lower the bond strength. A greater pulling force will provide more energy to the atoms, which speeds up the atomic motion along the force direction without sufficient time for relaxation. A weaker bond will break more easily under the same force. Therefore, at higher temperatures and under larger forces the Au wire will break swiftly, as illustrated in Figure 3a and Figure 3b.

In order to understand the fluctuation character in the intermediate force region of the curve in Figure 3b, we analyzed the breaking modes and structure transition of the Au wire during pulling, as shown in Figure 4 and the supplementary multimedia movies. The structure of the Au wire undergoes different structural transforming process when it is under different tensile forces $f$. When $f > 0.1$ nN/atom, the breaking and structure transformation are limited to a length scale nearby the pulling region, the breaking time depends mainly on the separating speed of the atoms in the pulling region from the main part of the wire. At $f <$ 0.1 nN/atom, a transition from BCT to FCC occurs near the pulling region and the transition diffuses into the neighboring region and the transition distance increases as the $f$ decreases. The structure transition prolongs the wire and then postpones the breaking time. In the simulation, we noticed the presence of a monatomic chain of 5-7 atoms under a weaker force (0.03 and 0.05 nN/atom for instances), which prolongs the breaking of the entire nanowire. However, the monatomic chain appears and breaks randomly, which is responsible for the fluctuation in the curve of Figure 3b. When $f < 0.018$ nN/atom, the tensile force is too weak to break the bond.

One may wonder how the strength of the tensile force could influence the structure of the Au nanowires. As derived from Eq. 1, the interatomic force acting on an atom depends on the distances between the neighbouring atoms, which suggest that the diffusion of the pulling effect depends on the directional motion of the atoms. Since an atom must move at a finite velocity, the pulling effect should diffuse along the wire in a limited speed, so does the structure transition. Strong tensile force will cause the atoms in the pulling region move more rapidly along the axis and increase the distance between the atoms over the edge of the pulling region in a short time, which causes the wire breaks quickly. So there is no enough time for the structure transformation diffusing, which causes the wire to break nearby the pulling region, without obviously structural transformation (Figure 4, $f$ = 1nN/atom). As the tensile force decreases, the slower motion of the atoms provides the nanowire enough time to transfer the pulling effect to other regions for relaxation, which results in the larger-scale structure transformation from BCT to FCC (Figure 4) and postpones the extension of chemical bond, then move the breaking point of the wire to the middle or near the fixed end. We noticed that when $f$ < 0.1 nN/atom, it is difficult to predict the position and time accurately when a narrow atomicwire appears.

## 4. Conclusion

It is found that the kinetics, the mode, and the lifetime of Au nanowire breaking are sensitive to the strength of the tensile force and the temperature of operating. The wire breaks abruptly at higher temperatures and under stronger forces. If the force is weak enough, no rupture occurs. In the intermediate force region, the wire breaks after a structure transition with the length scale of diffused transition being subject to the force applied. The similar effects might exist in other kinds of materials in nanoscale such as nanobelts and nanoclusters.

Figure Captions

Figure 1 Illustration of interatomic potential energy $E$, interactive force $F$, and the slope of $F$ ( $D = \partial F_{ij}/\partial r_{ij}$ ) between two Lennard-Jones atoms. The turning pint of the D curve corresponds to the breaking limit of the bond.

Figure 2 Illustration of the pulling scheme for the nanowire. The left end is fixed; the tensile force $f$ is exerted on each atom uniformly in the pulling region at the right end of the nanowire under various temperatures.

Figure 3 (a) Temperature and (b) tensile strength dependence of the breaking time of the Au nanowire under constant force for (a) and constant temperature for (b). The inset zooms in the breaking time at smaller forces showing random fluctuations.

Figure 4 The breaking modes of the Au nanowire when it is pulled under different tensile forces at 50 K.

Multimedia movie illustration: the breaking kinetics and modes under the forces of 0.1 and 0.04 nN/atom at 50 k.

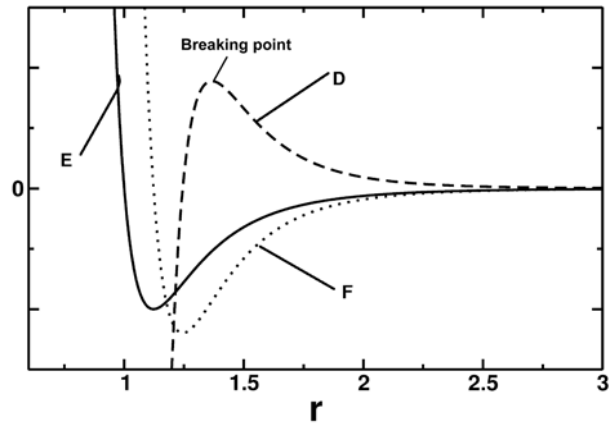

Fig 1

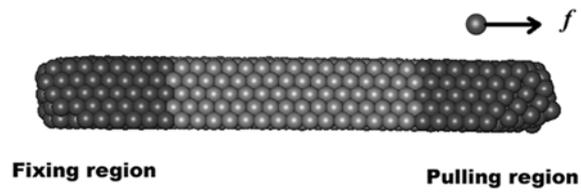

Fig 2

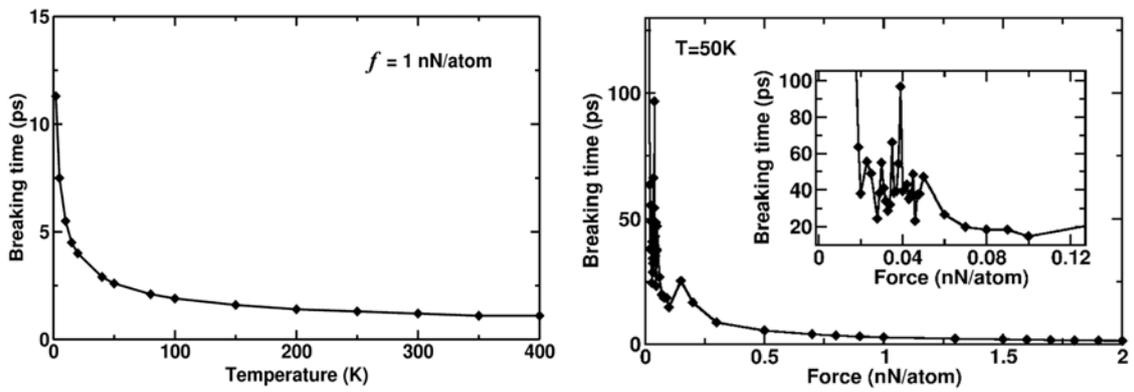

(a)   (b)

Fig 3

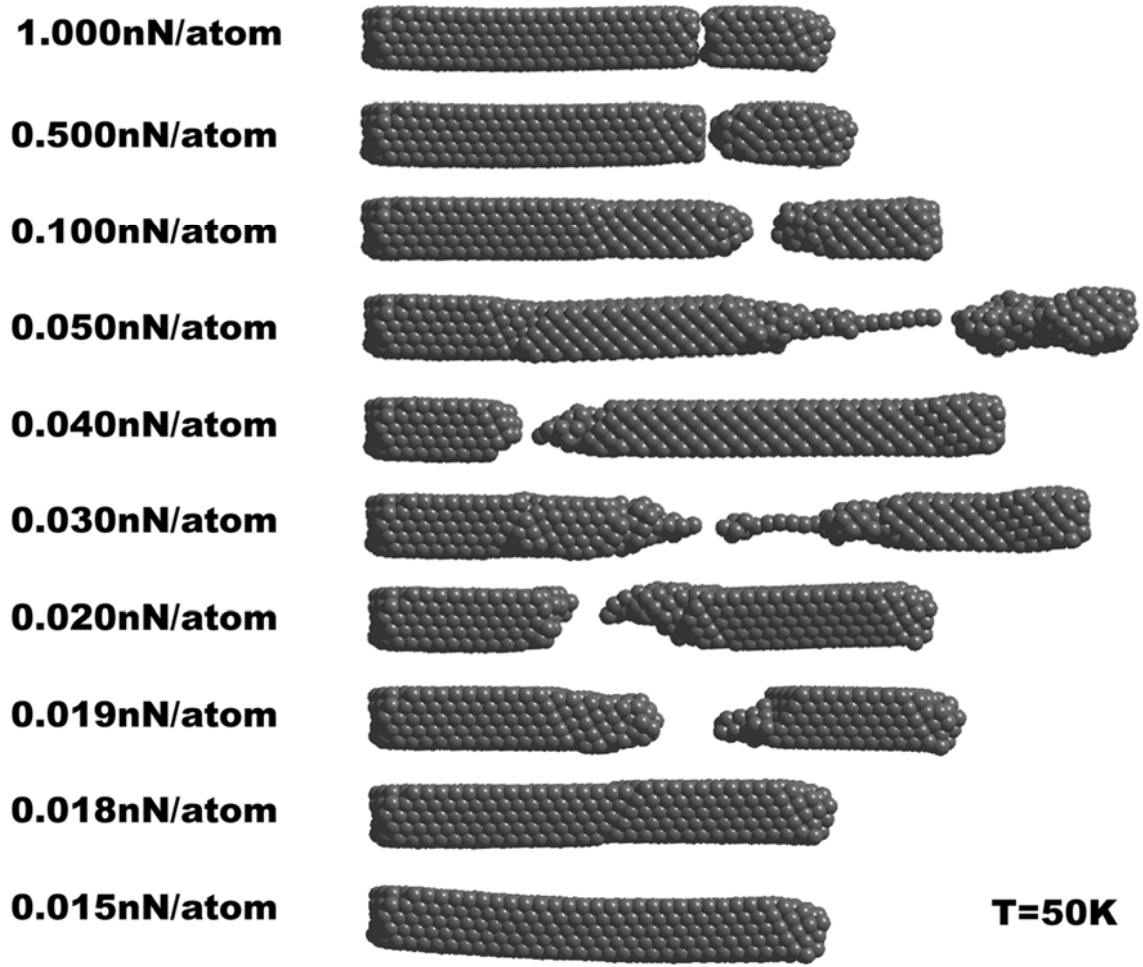

Fig 4